\documentclass[aps,pra,twocolumn,showpacs,superscriptaddress,tightenlines,preprintnumbers,sort&compass,amsmath,amssymb,floatfix,nofootinbib]{revtex4}

\newcommand\BER{\mathrm{QBER}}
\newcommand\ket[2]{|#1\rangle_{#2}}
\newcommand\bra[2]{{}_{#2}\langle#1|}
\newcommand\oprod[2]{\ket{#1}{#2}\bra{#1}{#2}}

\usepackage{epstopdf}
\usepackage{graphicx}
\usepackage{ulem}

\def\extra#1{{``#1''}}
\def \etal{\textit{et al.}}

\begin{document}

\title{Experimental {E}avesdropping {B}ased on {O}ptimal {Q}uantum {C}loning}

\author{Karol Bartkiewicz}
\email{bartkiewicz@jointlab.upol.cz} \affiliation{RCPTM, Joint
Laboratory of Optics of Palack\'y University and Institute of
Physics of Academy of Sciences of the Czech Republic, Faculty of
Science, Palack\'y University
\\17. listopadu 12, 771~46 Olomouc, Czech Republic}

\author{Karel Lemr}
\email{k.lemr@jointlab.upol.cz}
\affiliation{RCPTM, Joint Laboratory of Optics of Palack\'y
University and Institute of Physics of Academy of Sciences of the
Czech Republic, Faculty of Science, Palack\'y University
\\17. listopadu 12, 771~46 Olomouc, Czech Republic}

\author{Anton\'in \v{C}ernoch}
\affiliation{Institute of Physics of Academy of Science of the
Czech Republic, Joint Laboratory of Optics of PU and IP AS CR, 17.
listopadu 50A, 77207 Olomouc, Czech Republic}

\author{Jan Soubusta}
\affiliation{Institute of Physics of Academy of Science of the
Czech Republic, Joint Laboratory of Optics of PU and IP AS CR, 17.
listopadu 50A, 77207 Olomouc, Czech Republic}

\author{Adam Miranowicz}
\affiliation{Faculty of Physics, Adam Mickiewicz University,
PL-61-614 Pozna\'n, Poland}

\begin{abstract}
The security of quantum cryptography is guaranteed by the
no-cloning theorem, which implies that an eavesdropper copying
transmitted qubits in unknown states causes their disturbance.
Nevertheless, in real cryptographic systems some level of
disturbance has to be allowed to cover, e.g., transmission losses.
An eavesdropper can attack such systems by replacing a noisy
channel by a better one and by performing approximate cloning of
transmitted qubits which disturb them but below the noise level
assumed by legitimate users. We experimentally demonstrate such
symmetric individual eavesdropping on the quantum key distribution
protocols of Bennett and Brassard (BB84) and the trine-state
spherical code of Renes (R04) with two-level probes prepared using
a recently developed photonic multifunctional quantum cloner
[K. Lemr \etal, Phys. Rev. A~\textbf{85}, 050307(R) (2012)]. We demonstrated that
our optimal cloning device with high-success rate makes the
eavesdropping possible by hiding it in usual transmission losses.
We believe that this experiment can stimulate the quest for other
operational applications of quantum cloning.
\end{abstract}


\maketitle

During the last decades, there has been much interest in
secure quantum communication~\cite{Gisin02, Chen10}. Quantum
key distribution (QKD) devices (apart from quantum metrology,
random number generators and adiabatic computers based on
quantum annealing) are arguably the only second-generation
quantum technologies providing commercially available
applications of quantum information and quantum optics up to
date~\cite{Georgescu}. The security of QKD follows from
Heisenberg's uncertainty relation or, equivalently, the
no-cloning theorem. However, QKD can be secure only below
some level of noise that unavoidably occur in any physical
system. Therefore, security bounds of QKDs are expressed in
terms of tolerated losses or noise.

For QKD to be secure Alice and Bob must operate on single photons;
hence, they need a single-photon source (SPS). SPSs are usually
implemented as a weak coherent pulse of light~\cite{Gisin02}; thus,
QKD are prone to photon-number splitting attacks. This attack can
be circumvented by, e.g., using decoy states~\cite{Hwang03} or
heralded SPS instead of weak coherent pulses. Since there are no
lossless channels, if the eavesdropper (Eve) is equipped with a
proper cloning machine mimicking the lossy channel then she can
clone (a part of) the state sent by Alice, while hiding her
presence in usual transmission losses.

Recent proposals of applications of quantum cloning~\cite{cloning}
range from quantum cryptography~\cite{Gisin10} and quantum
metrology~\cite{qmetrology} to nonclassicality tests in
microscopic-macroscopic systems~\cite{Martini08} and, even,
proposals related to quantum experiments with human
eyes~\cite{Sekatski09}.

In this Letter, we experimentally demonstrate the usefulness
of cloning for quantum cryptoanalysis, i.e., for the
eavesdropping of QKD over noisy quantum channels.

There are a number of well-known QKDs including the famous
BB84 of Bennett and Brassard~\cite{BB84} based on mutually
unbiased bases and the biased-bases R04 of Renes~\cite{Renes04}.
Attacks on those protocols can be classified as individual (or
incoherent) and coherent (including joint and
collective)~\cite{Gisin02}. Every attack can be imagined as
follows: Eve sends a photon (probe) prepared in some polarization
state which interacts with a photon sent by Alice, then Eve sends
a photon to Bob and performs a measurement on her probe (she might
wait until the key sifting process is over). Recently, attacks on
QKD were proposed exploiting technological loopholes
rather than the limits imposed by
physics~\cite{Lydersen10,Gerhardt11}. However, in this Letter we analyze the
physical bounds on the security of the QKDs.

We focus only on individual attacks on the two QKDs assuming
that Eve waits until Alice and Bob complete key sifting and
then performs her measurements. This kind of attack requires
that Eve has access to quantum memory (QM) in order to store
her probes during the key sifting, but does not require Eve
to perform a coherent measurement on many photons at a time.
Such satisfactory memory has not been invented yet; however,
recent encouraging results~\cite{Specht11} carry the promise of
realizing good QMs in near future. Moreover, in our opinion,
coherent multi-qubit readout may require an additional
technological leap. For clarity of presentation we focus only
on trine-state R04 and BB84. Nevertheless, our approach can
be used for analyzing generalizations of those protocols. By
referring hereafter to R04 we mean its trine-state version.

It is known that the acceptable quantum bit error rate
($\BER$), i.e., the ratio of the number of wrong qubits to
the total number of qubits received, is
$15\%$~\cite{Lutkenhaus96} for BB84 and
$16.7\%$~\cite{Renes04} for R04 assuming an individual attack
with a four-level probe and that Eve does not wait for the
key sifting. These $\BER$ bounds could suggest that R04 is
more robust to eavesdropping than BB84. However, it was shown
that, by assuming one-way communication between Alice and
Bob, BB84 is unconditionally secure if $\BER\leq
11\%$~\cite{Shor00}, while R04 if $\BER\leq
9.85\%$~\cite{Boileau05}. On the other hand, if Eve waits for
the key sifting process to finish, BB84 is secure if
$\BER\leq 14.6\%$~\cite{Fuchs97} (or $15\%$~\cite{Gisin97}
for the two-level probe), while the corresponding $\BER$
bound for R04 is unknown to our knowledge. Nevertheless, in
this Letter we show that the $\BER$ bound for R04 and BB84 is
$16.7\%$ for the optimal cloning attack with a two-level
probe.

\begin{figure}
\includegraphics[width=8cm]{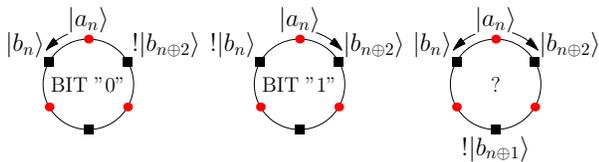}
\caption{\label{fig:r04}(Color online) Diagram describing
R04~\cite{Renes04}. Alice (Bob) publicly agrees beforehand to send
(measure) one of the trine states marked by red (black) dots,
respectively. Both agree that the clockwise (anticlockwise)
sequence of their states corresponds, e.g., to bit 1 (0). Bob
publicly informs Alice what he has \emph{not} measured (marked by
an exclamation mark). Alice ignores the inconclusive cases (and
informs Bob about them). In the other two cases, Alice and Bob
obtain the same bit value.}
\end{figure}

The algorithm for the cloning-based eavesdropping
investigated in our Letter reads as follows: (i) Eve plugs a
cloning machine together with QM into the quantum
communication channel between Alice and Bob. (ii) Alice sends
one of the states used in BB84 or R04. (iii) Eve intercepts
the state and prepares two noisy copies. This cloning
introduces losses. (iv) Eve sends one copy to Bob and keeps
the other copy in QM. (v) Bob measures the received copy.
(vi) Alice and Bob publicly perform key sifting. (vii) Eve
performs positive-valued measures (POVMs) on each of the
stored qubits to guess the bit value that was obtained by
Alice and Bob simultaneously. She assigns corresponding bit
values to her measurement outcomes. The steps performed by
Eve are discussed below (for additional details see
Ref.~\cite{Supplement}). Since, in our experiment, we do not
have access to QM we simulate it by performing a reconstruction
of the two-qubit density matrix shared by Eve and Bob and
later by projecting Bob's part of the state onto one of the
bases used in QKD. The reduced density matrix describing
Eve's qubit is assumed to be stored in QM.

\begin{figure}
\includegraphics[width=8cm]{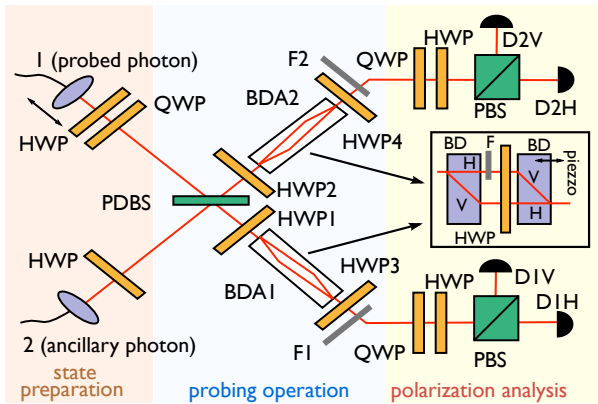} 
\caption{\label{fig:setup}(Color online) Experimental setup
as described in the text. States of the probed and ancillary
photons are prepared with half-wave (HWP) and quarter-wave
(QWP) plates. The photons overlap on the
polarization-dependent beam splitter (PDBS) and undergo
polarization-sensitive filtering in the beam divider
assemblies (BDAs). Each BDA (see figure inset) consists of a
pair of beam dividers (BDs), a neutral density filter (F), and
a half-wave plate (HWP). The tomography of the two-photon state
is accomplished by means of the HWPs, QWPs, polarizing beam
splitters (PBDs), and single-photon detectors (D).}
\end{figure}

Symmetric attacks on QKD can be performed by using a
multifunctional optimal quantum cloner (OQC)~\cite{Lemr12}.
In R04~\cite{Renes04} (explained in Fig.~\ref{fig:r04}) Alice
sends one of the three equally separated equatorial qubits
$|a_n\rangle = \mathcal{N}[|H\rangle +\exp(i2n\pi/3)|V\rangle
]$ and Bob detects $|b_n\rangle = \mathcal{N}[|H\rangle
+\exp(i2n\pi/3+i\pi/3)|V\rangle ]$, where $n=0,\,1,\,2$ and
$\mathcal{N}=1/\sqrt{2}$. Since all the states used in R04
(and also in BB84) are on the equator of the Bloch sphere
(say $xy$ plane), we require that Eve's action causes {the
Bloch sphere of the} qubit received by Bob to shrink
uniformly in the $xy$ plane {(qubit's purity decreases)} so
that her presence cannot be easily detected. Thus, the
density matrix of Bob's qubit reads {as} $\rho_B =
\frac{1}{2}\left[\openone +
(\hat\eta_B\vec{r}_B)\cdot\vec{\sigma}_B \right]$, where the
Bloch-sphere shrinking is described by matrix $\hat\eta_B =
\mathrm{diag}(\eta,\eta,\eta_\bot)$, where $\eta$
($\eta_\bot$) is the shrinking factor in the $xy$ plane ($z$
direction), $\vec{r}$ is the Bloch vector of the initial
qubit, and $\vec{\sigma} = (\sigma_x,\sigma_y,\sigma_z)$ is a
vector of Pauli's matrices. Our OQC~\cite{Lemr12} provides
the following shrinking factors $\eta =
2\sqrt{p}\Lambda\bar{\Lambda}$ and $\eta_\bot = \Lambda^2 +
\bar{\Lambda}^2(p-q)$, where $q+p=1$ and
$\Lambda^2+\bar{\Lambda}^2 = 1$ assuming that
$p,\,q,\,\Lambda,\bar{\Lambda}\in [0,1]$, where $p$ is the
asymmetry parameter of the clones and $\Lambda$ is the
cloning ``strength'' since it affects the purity of the
clones (related to the shrinking factors) in the same way. In
our experiment we fix values of $p$ and $\Lambda$ by
adjusting polarization sensitive filtering in BDAs (see
Fig.~\ref{fig:setup} and Ref.~\cite{Supplement}). Moreover,
for Eve's probe we obtain $\hat{\eta}_E(p,\Lambda) =
\hat{\eta}_B (q,\Lambda)$. This operation is similar to the
one of the mirror phase-covariant
cloner~\cite{Bartkiewicz09,Lemr12}. The difference depends on
$p$ which implies that the states of Eve and Bob have
different fidelities with respect to the states sent by
Alice. Furthermore, the fidelity of Bob's qubits is
$F_B(p,\Lambda) = (1+2\sqrt{p}\Lambda\bar{\Lambda})/2$,
whereas Eve obtains $F_E(p,\Lambda)=F_B(q,\Lambda)$. The
unitary cloning transformation reads as
$|H\rangle_{A}\to[\Lambda|H,H,0\rangle+\bar\Lambda|\psi(p),1\rangle]_{B,E,{\rm
anc}}$ and
$|V\rangle_{A}\to[\Lambda|V,V,1\rangle+\bar\Lambda|\psi(q),0\rangle]_{B,E,{\rm
anc}}$, where
$|\psi(p)\rangle=\sqrt{p}|H,V\rangle+\sqrt{q}|V,H\rangle$.
The resulting state shared by Bob and Eve is obtained by
tracing out the ancilla, which in our experiment corresponds
to random switching between $H$- and $V$-polarized photons
used by Eve as probes. For $p=\Lambda^2=1/2$ the OQC becomes
the symmetric $1\to2$ phase-covariant cloner~\cite{Bruss00},
which for BB84 causes $\BER=1-F_B=14.6\%$. Moreover, for
$p=1/2$ and $\Lambda^2=2/3$, the OQC becomes the universal
cloner~\cite{Buzek96}. We assume Eve's probe to be a qubit,
while the most general approach requires the probe to be a
four-level system. Our restriction is valid if two-photon
interactions~\cite{Lemr12} are only used for the
eavesdropping.

\textit{Optimal eavesdropping strategy. ---} Eve knows the
initial state of her photon as her OQC performs conditional
operations~\cite{Lemr12,Supplement}, where the asymmetry is
implemented by introducing additional losses~\cite{Bart07}.
However, Eve, to optimize her attack on R04, must choose the
optimal strategy for distinguishing between Bob's measurement
results $b_{n}$ and $b_{n\oplus1}$ given that Alice sent
$a_{n}$ and $a_{n\oplus 2}$, respectively, where $\oplus$
stands for sum modulo 3. While restricting Eve's readout to
the von Neumann's measurements we found the optimal ones
maximizing Eve's information (this follows from the symmetry
of the shrinking factors) to be equivalent to Helstrom's
measurements~\cite{Chefles00} discriminating between states
$|b_{n}\rangle$ and $|b_{n\oplus1}\rangle$ (or
$|a_{n}\rangle$ and $|a_{n\oplus 2}\rangle$) independent of
the values $\Lambda$, $p$, and the initial state of the
probe. Thus, Eve's measurement is a projection on equatorial
qubits of phase $2n\pi/3+\pi/6+m\pi$ ($2n\pi/3+5\pi/6+m\pi$)
if Bob's message (see Fig.~\ref{fig:r04}) is $!|b_{n}\rangle$
($!|b_{n\oplus 2}\rangle$), where $m=0,1$ is Eve's bit value.
For BB84, Eve uses the measurement as Bob.

\begin{figure}
\includegraphics[width=8cm]{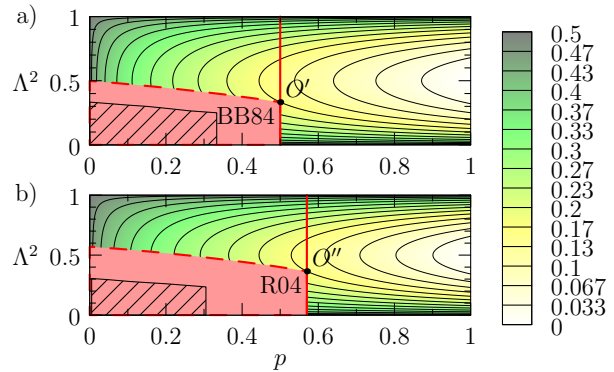}
\caption{\label{fig:QBER} (Color online) Cloning parameters,
$\BER$ and the cloning-attack security of the (a) BB84 and
(b) R04 as a function of the cloning asymmetry parameter $p$
and cloning ``strength'' $\Lambda$. Dashed red lines show the
$\BER$ bound corresponding to the zero-length distilled key,
i.e., $R=I_{A,B}-\min(I_{A,E},I_{B,E})=0$. Thus, cloning
enables successful eavesdropping in the regions marked 
by names of the QKDs. The optimal cloning attacks cause $\BER=16.7\%$ 
if $p=0.57$ and $\Lambda^2=0.36$ (point $O^{\prime\prime}$) for
R04 and $p=0.5$ and $\Lambda^2=0.33$ (point $O^{\prime}$) for
BB84. The vertical solid red lines show the $\BER$ bounds on the
privacy of directly transmitted information corresponding to
$I_{A,B} = I_{A,E}$, which are equal to $15.0\%$ for R04 and
$14.6\%$ for BB84. The area of $\Lambda^2\geq 1/2$
($\Lambda^2=1/2$) corresponds to the mirror phase-covariant
OQC~\cite{Bartkiewicz09} (the asymmetric phase-covariant
OQC~\cite{cloning,Bart07}).
Hatched areas indicate the range of the cloning attacks
without using quantum memory.}
\end{figure}

For the measurements we calculated~\cite{Supplement} the
mutual Shannon information $I_{X,Y}$ between the three users,
where $X,Y$ stand the initials of the corresponding parties.
Next, we calculated the secret-key rate (i.e., the lower
bound on the distilled key length per number of the
sifted-key bits)
$R=I_{A,B}-\min(I_{A,E},I_{B,E})$~\cite{CK78} as a function
of $\Lambda$ and $p$. Finally, we found the optimal cloning
attack by maximizing $I_{A,B}$ for $R=0$. The results of our
theoretical analysis, as summarized in Fig.~\ref{fig:QBER},
imply that the optimal (cloning restricted) two-level-probe
individual attack on R04 yields $\BER=16.7\%$ for
$\Lambda^2=4/11$ and $p=4/7$. Our results for the analogous
strategy for BB84 are shown in Fig.~\ref{fig:QBER}, where the
best attack yields $\BER=16.7\%$ for $\Lambda^2=1/3$ and
$p=1/2$. The $\BER$ depends on the fidelity of cloning
[$\BER=1-F_B$ for BB84 and $\BER=4(1-F_B)/(5-2F_B)$ for R04],
while the information extracted by Eve depends both on the
fidelity (as does $I_{{A,E}}$) and the entanglement of clones
(correlations between Bob's and Eve's qubits). Thus, the
optimal attack must balance these two quantities to provide
$R=0$ for a given $I_{A,B}$.

\textit{Experimental aspects of the eavesdropping. ---} In order to
implement the cloning attack, we employed the experimental setup
(see Fig.~\ref{fig:setup}), which consists of three main parts:
the source of photon pairs, the cloner and the two-photon
polarization analyzer. Spatially-separated photon pairs of
$\lambda=826$\,nm wavelength are created in noncolinear type I
degenerate spontaneous parametric down-conversion process in
LiIO$_3$ crystal (1\,cm thick) pumped by a cw Kr$^+$ laser beam
(TEM$_{00}$ mode, 250\,mW of optical power). Our source
approximates two synchronized SPSs with the accuracy adequate for
our demonstration, since the probability of having more than one
photon in a mode for the 1\,ns detection window is much lower than
the probability of single-photon detection (approx.
$10^-5$~\cite{Supplement}). The emitted photons are in a separable
polarization state; hence, Alice's state encoded as the signal does
not change the polarization of the probe. Random choice of the
states sent by Alice ensures that the polarization of the two
photons is uncorrelated. This corresponds to having two
independent but synchronized SPSs. However, in a real attack Eve
would have to use a separate SPS. The photons propagate from the
source to the OQC input via single-mode fibers. The photons are
coherently superposed on the polarization-dependent beam splitter
(PDBS). Next, the photons are subjected to polarization-sensitive
filtering in both output modes (see BDA in Fig.~\ref{fig:setup}).
Finally, we postselect on coincidences -- one photon in each of
the two output modes of the cloner -- and carry out polarization
analysis of the two-photon state~\cite{Halenkova12}. Using our
tomographical data, we estimated the two-photon density matrix
applying the maximum likelihood method~\cite{Jezek03estimation}.
We used the tomography results to numerically simulate Eve's
attack assuming that she probes Alice's photon, keeps the probe
until key sifting, and passes the probed photon to Bob (for
details see Ref.~\cite{Supplement}). We calculated the $\BER$ and
secret-key rate and compared them with theoretical predictions in
Table~\ref{tab:exp}.

\begin{table}
\caption{\label{tab:exp} Performance of the OQC for BB84 and R04.
The experimental values (subscript $E$) of the $\BER$ and the
secret-key rate $R$ calculated from the measured density matrices
are compared with theoretical predictions (subscript $T$). The
success probability $p_s$ of the OQC was estimated as in
Ref.~\cite{Lemr12}. The OQC parameters $p$ and $\Lambda$ determine
the shrinking of the Bloch sphere due to the cloning.}
\begin{ruledtabular}
\begin{tabular}{llllllll}
QKD                          &  $R$       & $\BER$    & $p_s$
& $p$  & $\Lambda^2$            \\ \hline
$\mathrm{BB84}_T$  & 0.00      &  16.7\%   &  13.7\%      &  1/2                  &  1/3        \\
$\mathrm{BB84}_E$  & $0.03 \pm 0.03$      & $18.5\% \pm 1.5\%$      & $15.1\% \pm 1.1\%$      &  1/2                  & 1/3       \\
$\mathrm{R04}_T$   & 0.00       & 16.7\%     &  12.7\%    &  4/7                  &  4/11       \\
$\mathrm{R04}_E$    & $0.01 \pm 0.08$      & $18.0\% \pm 3.5\%$     &  $7.4\% \pm 0.1\%$     &   4/7                 &  4/11       \\
\end{tabular}
\end{ruledtabular}
\end{table}

The results indicate that our attack would be possible if QM was
available. However, to deploy this device in a real QKD network,
one has to consider several technological aspects of this attack.
First, because of its probabilistic nature, the OQC introduces
losses. The success probability of $10\%-20\%$ corresponds to
7-10\,dB losses. Observing such losses might indicate that the
line is insecure. Thus, Eve must mask these losses as usual
channel losses. Supposing typical fibers losses of 3.5\,dB/km (as
for the fibers in our experiment and in~\cite{Gisin02}), Eve would
need to replace 2-3\,km of the line with a fiber of negligible
losses. Using photons at telecom wavelengths would be more
practical than 826\,nm light for communicating over large
distances since, for the telecom-window wavelengths, the loses are
$\sim$0.2\,dB/km~\cite{Gisin02,Chen10}, which makes the distances
about ten times larger and Eve's task more difficult. Typical
detectors, designed for the telecom regime, provide low efficiency
of about 0.25 (approx. 6\,dB of losses) and high dark-count rate,
i.e., noise. Much larger losses, which could enable
eavesdropping, appear for long-range free-space transmission
reaching, e.g., 157\,dB for photons reflected from the Ajisai
satellite~\cite{Villoresi08}).

Eve should also control the unsuccessful cases when the signal and
the probe propagate to Bob, who can detect them and raise alarm.
Eve can achieve this by using quantum nondemolition (QND)
measurement~\cite{Bula13}. If she does not find a photon in her
output mode, she will close the line towards Bob. Finally, Eve's
attack relies on the perfect overlap between the signal and the probe
photons (Hong-Ou-Mandel's interference). Typical full width at
half maximum (FWHM) of photons generated via spontaneous
parametric down-conversion corresponds to tens of $\mu$m. Thus,
the requirement on the two-photon overlap is of the order of
$\mu$m. This corresponds to a few fs. Any jitter caused by Alice
leads to the reduced two-photon overlap, lower purity and fidelity
of the output states. Eve can however overcome this by performing
the QND detection at her OQC input, which triggers the cloning
process but requires photon generation on demand. In a real
cloning attack Eve has to prepare photons of the same spectral
properties as the photons sent by Alice. In our experiment we use
photons at 826\,nm with spectral width FWHM=8.9\,nm (160\,fs
coherence time). These are typical values reached by Alice using a
femtosecond laser as a photon source. Let us note that the photon
peaks need to overlap as perfectly as possible making the
acceptable time difference corresponding to a fraction of
coherence time which changes as wavelength squared divided by
FWHM. Hence, overlapping is easier for longer wavelengths (e.g.,
in the telecom window) and narrow FWHM. Both the parameters
should be tuned by Alice to maximize the security of the QKD.

\textit{Conclusions. ---} We investigated the feasibility of
symmetric individual attacks on BB84~\cite{BB84} and
R04~\cite{Renes04} assuming that Eve tracks the key sifting
and uses a multifunctional OQC~\cite{Lemr12}. We optimized
quantum cloning such that the minimum mutual information
between an eavesdropper and a legitimate user was equal to
the mutual information between the legitimate users at the
lowest $\BER$. Thus, legitimate users cannot distil a secret
key from their raw key bits. Consequently we found tolerable
$\BER$ for this kind of attack to be $16.7\%$ for BB84 and
R04. We performed the proof-of-principle experiment in which
$R\approx0$ was attained for $\BER=18.5\%\pm1.5\%$ for BB84 and
$\BER=18.0\%\pm3.5\%$ for R04. Our experiment together with the
reported progress in development of QM (see,
e.g.,~\cite{Specht11}) suggest that, even in the presence of
SPSs and perfect detectors, the QKD could be successfully
attacked with a probe similar to ours if Alice and Bob
tolerate too high $\BER$ (see Table~\ref{tab:exp}) or losses
(approx. $7$\,dB for our device). Our experiment shows that
the OQCs are interesting both from the fundamental and
practical points of view as tools of quantum cryptanalysis as
they establish the security bound for an important class of
QKDs.

We thank Ravindra Chhajlany, Anirban
Pathak, and Radim Filip for discussions. K.B. and A.M. were
supported by Grant No. DEC-2011/03/B/ST2/01903 
and No. DEC-2011/02/A/ST2/00305 of the Polish
National Science Centre. K.B. and K.L. were supported by Grants
No. CZ.1.05/2.1.00/03.0058, No. CZ.1.07/ 2.3.00/20.0017,
CZ.1.07/2.3.00/20.0058, CZ.1.07/2.3.00/30.0004
and No. CZ.1.07/2.3.00/30.0041, A.\v{C}. and J.S.
acknowledge support by the GA\v{C}R Grant No. P205/ 12/0382.


\title{Experimental {E}avesdropping  {B}ased on {O}ptimal {Q}uantum {C}loning:\\ {S}upplementary {M}aterial}

\clearpage
\begin{center}
{\Large Supplementary material:}
\end{center}

In this supplement we give more technical details on cloning
transformations, cloning-based hacking and experimental 
data processing. We also compare graphically our theoretical 
and experimental tomographic results.

\section{The cloning machine}
\subsection{Cloning transformation}
The general phase-covariant cloning transformation is given
by
\begin{eqnarray}
\ket{0}{a}&\to&\Lambda\ket{000}{b,e,c} + \bar\Lambda(\sqrt{p}\ket{01}{b,e} + \sqrt{q}\ket{10}{b,e})\ket{1}{c},\nonumber\\
\ket{1}{a}&\to&\Lambda\ket{111}{b,e,c} +
\bar\Lambda(\sqrt{p}\ket{10}{b,e} +
\sqrt{q}\ket{01}{b,e})\ket{0}{c}\label{eq:cloning}
\end{eqnarray}
in the Hilbert space of the two clones of Bob in mode $b$ and
Eve in mode $e$ extended by an ancillary mode~$c$. Moreover,
0 (1) denotes horizontal $H$ (vertical $V$) polarization,
while $\bar{\Lambda}=\sqrt{1-\Lambda^2}$ and $p=1-q$ are
positive real numbers. The mode $c$ has to be traced out to
provide the state shared by Bob and Eve. The
transformation~(\ref{eq:cloning}) equally disturbs all the
equatorial qubits,
\begin{equation}
\ket{a_n}{}=\frac{1}{\sqrt{2}}\left[\ket{0}{}+\exp\left(i\phi_n\right)\ket{1}{}\right],
\end{equation}
where $n=0,1,...$. We experimentally implemented the cloning
transformation, given by Eq.~(\ref{eq:cloning}), by fixing
the values of $\Lambda$ and $p$, and by introducing
polarization sensitive losses in  the BDAs.

Eve, to make her cloning-based eavesdropping not easily
detectable, ensures that 0's and 1's are measured by Bob with
the same probability. To achieve this, Eve uses one of the
following transformations with probability $1/2:$
\begin{eqnarray}
U_0\ket{a_n,0}{}&=&C_{0}[\Lambda\ket{00}{}+\bar\Lambda\mathrm{e}^{i\phi_{n}}(\sqrt{q}\ket{01}{} + \sqrt{p}\ket{10}{})],\nonumber\\
U_1\ket{a_n,1}{}&=&C_{1}[\Lambda\ket{11}{}
+\bar\Lambda\mathrm{e}^{i\phi_{n}}(\sqrt{q}\ket{10}{} +
\sqrt{p}\ket{01}{})],
\end{eqnarray}
where $C_{x}=\sqrt{p_{x}/2}$ with $p_{x}$ ($x=0,1$) being the
success probability of the transformation $U_x$ including the
transmission losses and the probability of having a single photon
in each mode. Note that $p_{x}$ should not depend on $n$ since the
cloning is phase-covariant, but because of experimental
imperfections this is not always true. However, if $p_x$ does not
change much with $n$m,  which is the case of our cloner
\cite{Lemr12}, one can replace it with the average over $n$. The
transformations $U_x$ are implemented by overlapping the signal
and probe photons on the polarization-dependent beam splitter and
polarization-dependent filters. It is apparent that in general the
resulting state shared by Bob and Eve is a mixed state. For
convenience in the following sections we will use shorthand
notation $\ket{\psi_{x,n}}{}=U_x\ket{a_n,x}{}$.

\subsection{Cloning parameters and polarization-sensitive losses}

In our implementation we randomly swap between $H$ and $V$
initial polarizations of the probe. If the probing photon is
initially $H$-polarized then in order to set $\Lambda$ and
$p$ we have to ensure the following transmittance ratios:
\begin{eqnarray}
\frac{\tau_{b,H}}{\tau_{b,V}}&=&\left( \frac{\Lambda}{\bar{\Lambda}\sqrt{p}}\right)^2\frac{(1-\mu)(1-\nu)}{(1-2\mu)^2},\nonumber\\
\frac{\tau_{e,H}}{\tau_{e,V}}&=&\left(
\frac{\Lambda}{\bar{\Lambda}\sqrt{1-p}}\right)^2\frac{\mu\nu}{(1-2\mu)^2},
\end{eqnarray}
where $\tau_{x,y}$ stands for the intensity transmittance of the
polarization $y=H,V$ in the spatial mode of Bob ($x=b$) and Eve
($x=e$); $\mu$ and $\nu$ stand for the amplitude transmittance for
the $H$ and $V$ polarizations of the polarization-dependent beam
splitter. In the ideal case, the latter transmittances should be
equal to  $\mu = \left(1+ 1/\sqrt{3}\right)/2$ and $\nu = \left(1-
1/\sqrt{3}\right)/2$. Due to manufacturing imperfections, the real
transmittances are $\mu = 0.77$ and $\nu = 0.19$. If the probe is
initially $V$-polarized, we fix
\begin{eqnarray}
\frac{\tau_{b,H}}{\tau_{b,V}}&=&\left( \frac{\bar{\Lambda}\sqrt{1-p}}{\Lambda}\right)^2\frac{(2\nu-1)^2}{2(1-\mu)(1-\nu)},\nonumber\\
\frac{\tau_{e,H}}{\tau_{e,V}}&=&\left(
\frac{\bar{\Lambda}\sqrt{p}}{\Lambda}\right)^2\frac{(2\nu-1)^2}{\mu\nu}.
\label{eq:tau}
\end{eqnarray}
All the transmittances are set according to Eq.~(\ref{eq:tau}) in
the corresponding BDAs.

\begin{figure}
\includegraphics[width=8cm]{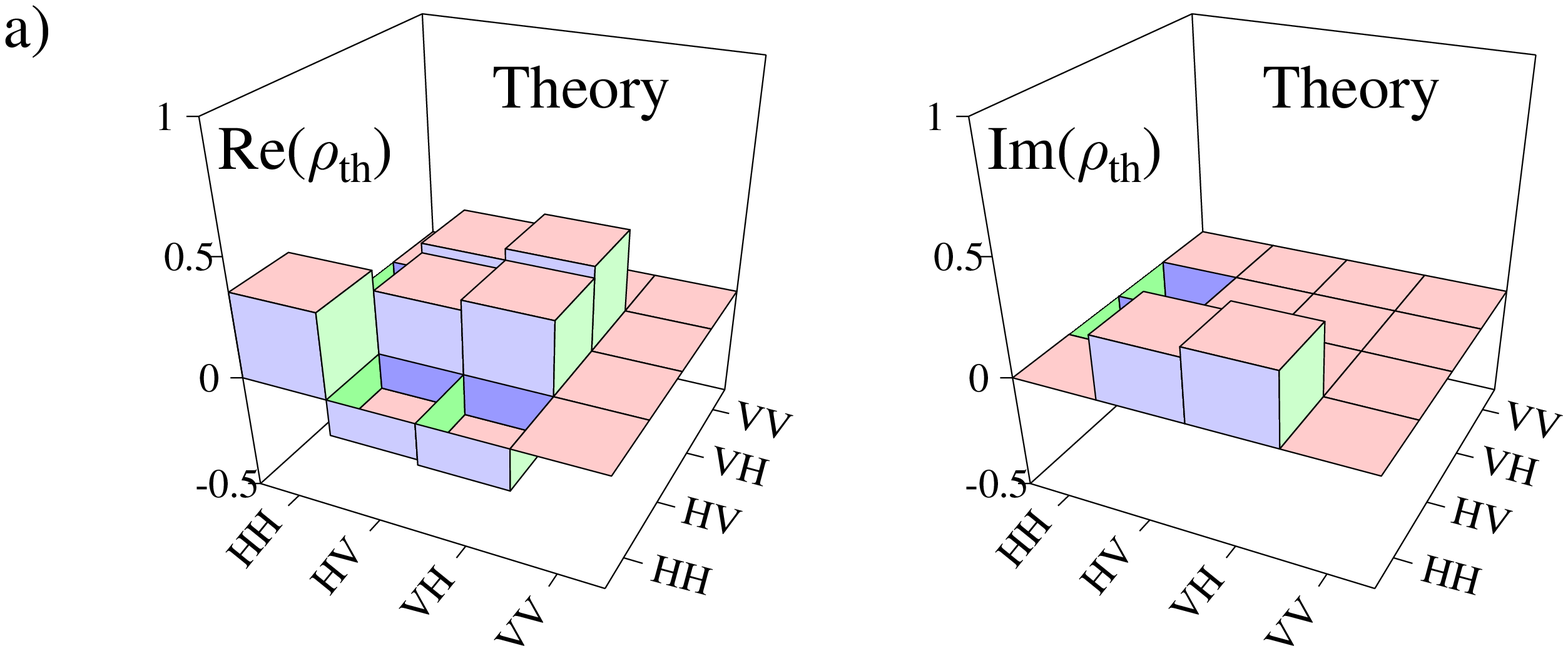}
\includegraphics[width=8cm]{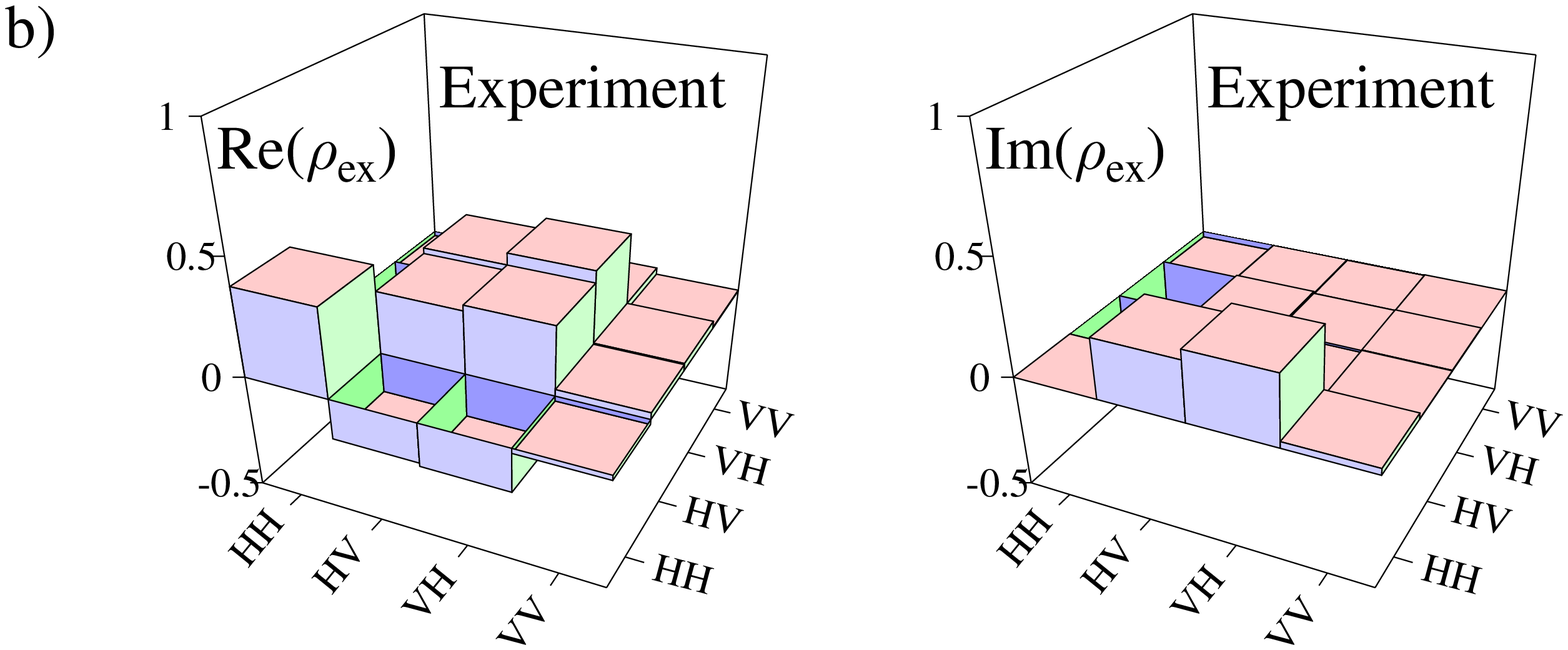}
\caption{\label{fig:tom1}(Color online) The reconstructed
two-photon state $\rho_{\mathrm{ex}}$ shared by Eve and Bob
compared to the experimental prediction $\rho_{\mathrm{th}}$
if Eve has probed $|a_1\rangle$ with the $H$-polarized photon
in R04: (a)~theory and (b)~experiment. The fidelity of
experimental matrix calculated as  $F =
\left(\mathrm{Tr}\sqrt{\sqrt{\rho_{\mathrm{th}}}
\rho_{\mathrm{ex}}\sqrt{\rho_{\mathrm{th}}}}\right)^2$ is
98\%. The complete density matrix describing Bob's and Eve's
photons is a mixture of density matrices for Eve using $V$-
and $H$-polarized photons.}
\end{figure}

\begin{figure}
\includegraphics[width=8cm]{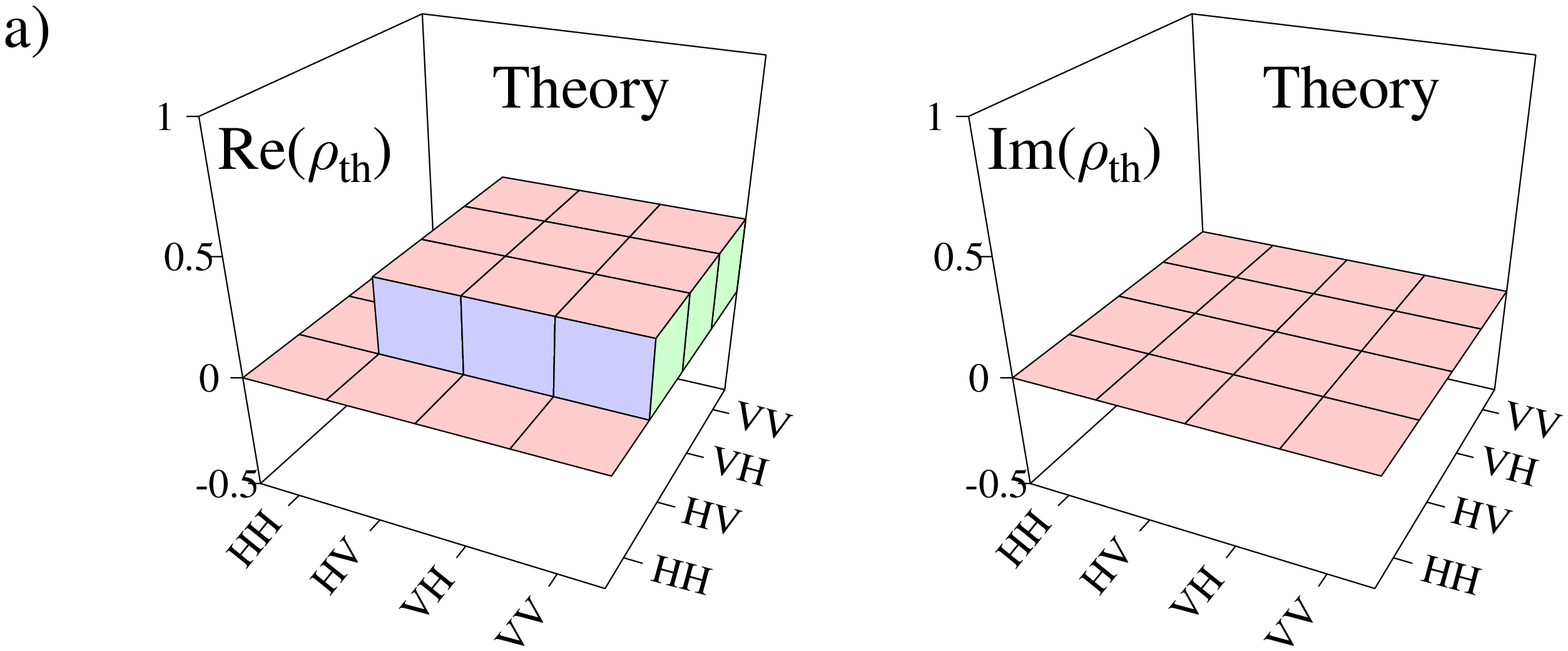}
\includegraphics[width=8cm]{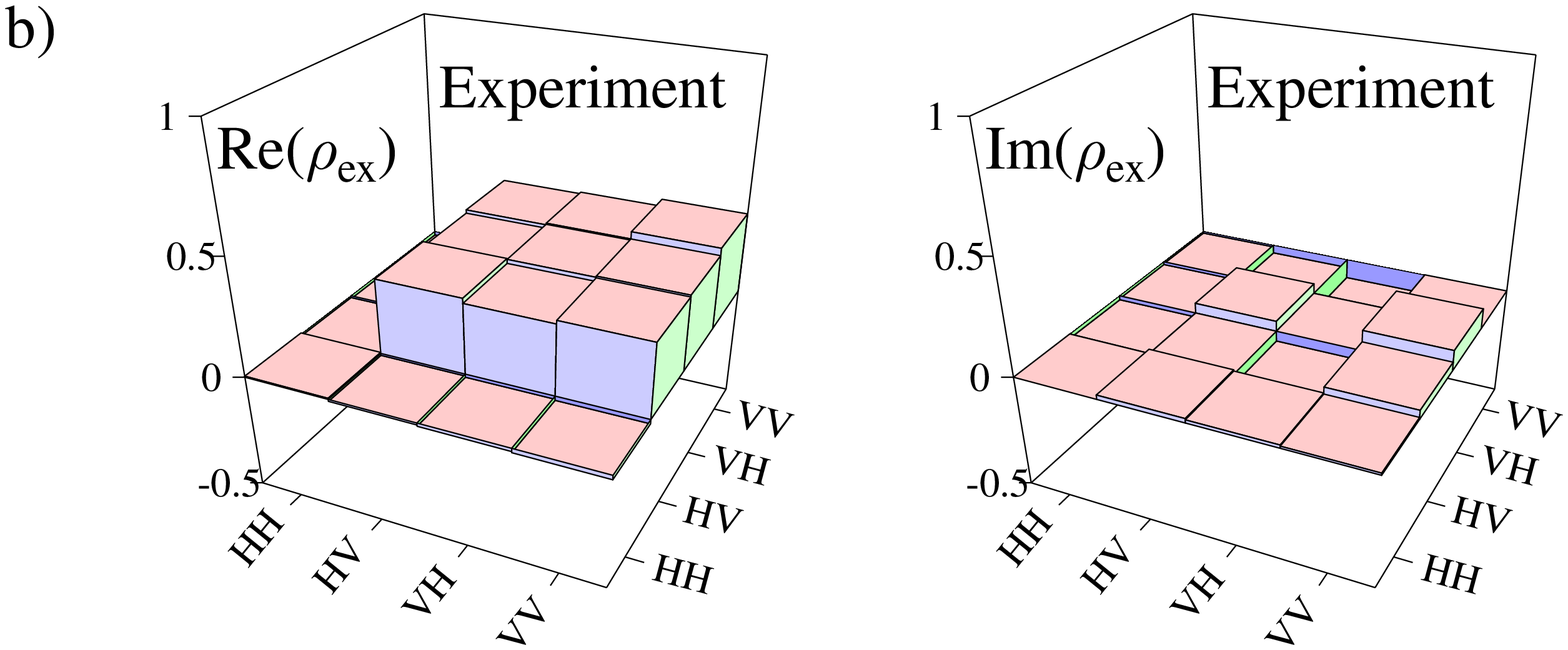}
\caption{\label{fig:tom2}(Color online) Same as in
Fig.~\ref{fig:tom1} but for BB84, where Eve has probed
$|a_0\rangle$ with the $V$-polarized photon. Here, the
fidelity of experimental matrix is 96\%.}
\end{figure}

\section{Gaining information from a clone}
\subsection{The R04 protocol}
In the R04 protocol states sent by Alice and then measured by Bob
read as
\begin{eqnarray}
\ket{a_n}{}&=&\frac{1}{\sqrt{2}}\left[\ket{0}{}+\exp\left(i\pi\frac{2n}{3}\right)\ket{1}{}\right],\nonumber\\
\ket{b_n}{}&=&\frac{1}{\sqrt{2}}\left[\ket{0}{}+\exp\left(i\pi\frac{2n+1}{3}\right)\ket{1}{}\right],
\end{eqnarray}
where $n=0,1,2$. Let us assume that Bob announces  that he did not
measure state $\ket{b_n}{}$. If the protocol is continued that
means that Alice sent $\ket{a_{n}}{}$ or $\ket{a_{n\oplus1}}{}$,
which corresponds to Alice obtaining bit value 1 or 0,
correspondingly. The bit value obtained by Bob depends on what he
measured. If Bob measured $\ket{b_{n\oplus2}}{}$
($\ket{b_{n\oplus1}}{}$), then his bit is 1 (0). Since Eve waits
for the Bob's announcement, she knows the number $n$. Moreover,
she knows the initial state of her qubit $\ket{x}{}$ and by using
this knowledge she performs the adequate POVM on her qubit. Eve
designs POVMs $\oprod{e_1(x,n)}{}$ ($\oprod{e_0(x,n)}{}$) in
advance in order to discriminate between Alice and Bob both
getting 1's (0's). The joint probability of obtaining bit values
$k$ by Alice, $l$ by Bob, and $m$ by Eve is given by
\begin{equation}
p_{x,n}(k,l,m)=N_{x,n}|\bra{b_{n\oplus(1+l)},e_m(x,n)}{}\psi_{x,n\oplus(1-k)}\rangle|^2,
\end{equation}
where the normalization constant $N_{x,n}$ ensures that
$\sum_{k,l,m} p_{x,n}(k,l,m)=1$. Therefore, the final
probability distribution is given as
\begin{equation}\label{eq:pR04}
p(k,l,m)=\frac{1}{6}\sum_{x=0}^1\sum_{n=0}^{2}p_{x,n}(k,l,m).
\end{equation}
In order to calculate $\ket{e_m(x,n)}{}$ let us first introduce
the following ancillary states
\begin{eqnarray}
\ket{\epsilon_m(x,n)}{}&=&\bra{b_{n\oplus
(1+m)}}{}\psi_{x,n\oplus (1-m)}\rangle,
\end{eqnarray}
where $m=0,\,1$. For the ancillary states we can calculate the
Bloch vectors as\begin{eqnarray} \vec{\epsilon}_m(x,n) =
\bra{\epsilon_m(x,n)}{}  \vec{\sigma} \ket{\epsilon_m(x,n)}{},
\end{eqnarray}
where $\vec{\sigma}=(\sigma_x,\sigma_y,\sigma_z)$. Then, one can
easily derive the Bloch vectors describing the Eve's POVMs as
follows
\begin{eqnarray}
\vec{e}_m &=& \frac{\vec{\epsilon}_m - \vec{\epsilon}_{1-m}}{|\vec{\epsilon}_m - \vec{\epsilon}_{1-m}|}\nonumber\\
&=&
(\sin\theta_m\cos\varphi_m,\sin\theta_m\sin\varphi_m,\cos\theta_m).\label{eq:vec}
\end{eqnarray}
These measurement directions are known to be optimal for the state
estimation and correspond to Helstrom's measurement. With the help
of Eq.~(\ref{eq:vec}), Eve's POVMs are defined by the following
pure states
\begin{equation}\label{eq:povm}
\ket{e_m}{} = \cos\frac{\theta_m}{2}\ket{0}{} +
\mathrm{e}^{i\varphi_m}\sin\frac{\theta_m}{2}\ket{1}{}.
\end{equation}
Therefore, for the R04 protocol we are able to easily calculate
the joint probability distribution of all the parties obtaining
various bit values by knowing the state shared by Eve and Bob
after the cloning transformation. In our experiment this state is
obtained by the two-qubit polarization tomography (see
Fig.~\ref{fig:tom1}).

\subsection{The BB84 protocol}

In the BB84 protocol states sent by Alice and measured by Bob read
as
\begin{eqnarray}
\ket{a_n}{}&=&\frac{1}{\sqrt{2}}\left[\ket{0}{}+\exp\left(i\pi\frac{n}{2}\right)\ket{1}{}\right],\nonumber\\
\ket{b_n}{}&=&\frac{1}{\sqrt{2}}\left[\ket{0}{}+\exp\left(i\pi\frac{n}{2}\right)\ket{1}{}\right],
\end{eqnarray}
where $n=0,1,2,3$. In case of the BB84 protocol Alice sends
$\ket{a_0}{}$ and $\ket{a_2}{}$ for the first basis (e.g., $X$
basis), and $\ket{a_1}{}$ and $\ket{a_3}{}$ for the second basis
(e.g., $Y$ basis). For the analyzed class of attacks, we assume
that Eve knows the basis because she listens to the public
announcements of Alice and Bob. Since Eve focuses only on the
cases when the bases of Alice and Bob match, we can simply write
the joint probability function as
\begin{eqnarray}
p_{x,y}(k,l,m)=N_{x,y}|\bra{b_{2l+y},e_m}{}\psi_{x,2k+y}\rangle|^2,
\end{eqnarray}
where $N_{x,y}=1/\sum_{k,l,m} p_{x,y}(k,l,m)$  and $x,y = 0,1$
enumerate the initial state of Eve's qubit and the basis in the
BB84 protocol, correspondingly. Therefore, the final probability
distribution, which is used to calculate the mutual information
between all the parties, is given as
\begin{equation}\label{eq:pBB84}
p(k,l,m)= \frac{1}{4}\sum_{x,y=0}^1 p_{x,y}(k,l,m).
\end{equation}
However, as in the case of the R04 protocol, we need to
express $\ket{e_m(x,y)}{}$ in terms of the information
available to Eve. The qubits obtained by Eve read as
\begin{equation}
\ket{\epsilon_m(x,y)}{}=\bra{b_{y+2m}}{}\psi_{x,y+2m}\rangle.
\end{equation}
Next, Eve constructs her POVMs analogously to the case of R04,
i.e., by first using Eq.~(\ref{eq:vec}) and then
Eq.~(\ref{eq:povm}).

\subsection{Experimental data processing}
Since, in our experiment, we automatically reconstruct a
two-photon density matrix describing a pure state shared by Bob
and Eve
\begin{equation}
\rho_{x,n}= p_x^{-1}\oprod{\psi_{x,n}}{}.
\end{equation}
In our experiment the purity of the two reconstructed components
is usually about 97\% (see Figs.~\ref{fig:tom1}
and~\ref{fig:tom2}). We are able to calculate the probability
distributions $p_{k,l,m}$ directly from the experimental data
using Eqs.~(\ref{eq:pR04}) and~(\ref{eq:pBB84}). This is because
we can rewrite the expressions for the probabilities comprising
the joint distributions using the  density matrix formalism and
the reconstructed matrices directly. Note that the expression for
the joint probabilities implicitly assume that the success rate is
uniform for all the configurations.

This success rate $p_s$, assumed to be uniform, is estimated in
the first approximation by $p_x$ averaged over all the equatorial
states used in the discussed protocols. In the experiment we
obtain $p_x$ by comparing coincidence rate with and without
applying the polarization-dependent beam splitter and
polarization-dependent filters. Moreover, if Eve wants Bob's
results to be symmetric with respect to swapping between the $H$
and $V$ polarizations, she should provide (if necessary by
introducing additional losses) $p_s=p_{x=0} = p_{x=1}$.
Alternatively, if $p_x$ varies little with $x$, Eve can do nothing
and then $p_s$ can be approximated by $(p_0+p_1)/2$. The success
rates $p_s$ given in Tab.~I were obtained by the following
prescription.

The measured probability distributions $p(k,l,m)\equiv
p_{A,B,E}(k,l,m)$ for Alice, Bob, and Eve having various bit
values are used to establish  the real quantum bit error rate
(QBER) for  the both analyzed protocols as discussed in  our
Letter. However, for establishing the secret bit rate $R$ of the
sifted key bits, first we determine the mutual Shannon information
between the parties using the standard formula
\begin{equation}
\label{Eq:mutinf}
I_{X,Y}=I_{Y,X}=\sum_{x,y=0}^{1}{p}_{X,Y}(x,y)\log_{2}\left[
\frac{{{p}_{X,Y}(x,y)}}{{p}_{X}(x){p}_{Y}(y)}\right],
\end{equation}
where  $p_{X,Y}(x,y)=\sum_{z=0}^1 p_{X,Y,Z}(x,y,z)$ and
$p_{X}(x)=\sum_{y=0}^1 p_{X,Y}(x,y)$ for $\{X,Y,Z\}=\{A,B,E\}$.
Using Eq.~(\ref{Eq:mutinf}) one can easily verify that the mutual
information between Alice and Bob is simply given by $I_{A,B}=1+
\delta\log_{2}\delta+(1-\delta)\log_{2}(1-\delta)$ in terms of the
probability $\delta$ of Bob obtaining a wrong bit value, which
corresponds to the QBER.

Note that the cloning operation works with the success rate $p_s <
1$ and for the remaining cases we do not detect any coincidences.
Thus, for the particular initial state of the cloning machine, a
fraction of $r =1-p_s$ of the photons  sent by Alice is absorbed
by the cloning machine. In our experiment this fraction is about
$r\approx80\%$, which corresponds to about 7\,dB of losses.

\section{Source of single photons}

In this section we demonstrate that the source of photons used in
our experiments approximates a single-photon source (SPS) to
be used by Alice in QKD. The SPS is approximated by one
of the modes of the down-converted light. In our
experiment the intensity of the down-converted light is low to
ensure that the probability of generating four (and larger
numbers) of photons is negligible. The photon statistics of such
down-converted beam is described by a thermal distribution
$P(n) = \bar{n}^n/(1+\bar{n}^{n+1})$~\cite{Gerry06} with a  small
average number of photons $\bar{n} \ll 10^{-3}$ (we ensure that in
the experiment). Thus, $P(n>1)\approx 0$ and we work with an
approximate source of photon pairs $|1\rangle|1\rangle$, where
each of the photons is in a different spatial mode.

Generating a photon in the first-order process is a random
event. The down-conversion events appear independently
of the number of down-converted photons $n$ in the
time window of 1\,ns, thus the probability 
of having $n$ photons is given by Poisson's distribution:
\begin{equation}\label{eq:p0a}
P(\lambda,n) = \lambda^n\exp(-\lambda)/n!,
\end{equation}
where $\lambda=\bar{n}$ is the average number of photons in the
time window. We can calculate the probability of not detecting any
photons in the 1\,ns window directly from the number of
detector clicks $N=35 \times 10^3$  that appeared as $P_0 = 1
- N\times10^{-9} = 0.999965$. This is correct if the usual time
interval between two detection events is larger than the dead time
of the detectors corresponding to about 35\,ns. On the other hand,
$P_0$ can be also calculated using a mathematical model of the
ON/OFF  (bucket) detectors of efficiency $\eta$ (in our experiment
$\eta = 0.5$) as
\begin{equation}\label{eq:p0b}
P_0 = \sum_{n=0}^{\infty}P(\lambda,n)(1-\eta)^n,
\end{equation}
where the only unknown $\lambda$ is estimated numerically to be
$\lambda = 7.00\times 10^{-5}$ assuming $P_0 = 0.999965$ and
$\eta = 0.5$. Therefore the probability of having more than one
photon in the time window is equal to
\begin{equation}
P_{>1}=1-\sum_{n=0}^1 P(\lambda=7\times 10^{-5},n)=2.45\times 10^{-9}.
\end{equation}
Moreover, since the probability of not having photons in the
35\,ns dead time window is very high as it reaches
$P^{35}(\lambda=7.00\times 10^{-5},n=0)=0.9976$, the
above-presented reasoning provides results in good agreement
with our other estimates.


\end{document}